\documentclass[aps,amsmath,amssymb,nofootinbib,twocolumn,prl]{revtex4-2}
\usepackage{color}
\usepackage[colorlinks=true,linkcolor=blue,urlcolor=blue,citecolor=blue]{hyperref}
\usepackage{color}

\usepackage{times}

\usepackage{graphicx}

\usepackage{tikz}
\usetikzlibrary{arrows,shapes,positioning,decorations.pathmorphing}
\usetikzlibrary{decorations.markings}
\usetikzlibrary{snakes}
\tikzset{snake it/.style={decorate, decoration=snake,segment length=3mm}}
\tikzstyle arrowstyle=[scale=1]
\tikzstyle directed=[postaction={decorate,decoration={markings,
    mark=at position .65 with {\arrow[arrowstyle]{stealth}}}}]
\tikzstyle endreversedirected=[postaction={decorate,decoration={markings,
    mark=at position 1.0 with {\arrow[arrowstyle]{stealth}}}}]
\tikzstyle enddirected=[postaction={decorate,decoration={markings,
    mark=at position 1.0 with {\arrow[arrowstyle]{stealth}}}}]
\tikzstyle reverse directed=[postaction={decorate,decoration={markings,
    mark=at position .65 with {\arrowreversed[arrowstyle]{stealth};}}}]

\usepackage{graphicx}

\newcommand{\Fig}[1]{\includegraphics[width=\columnwidth]{./#1}}
\renewcommand{\section}[1]{\smallskip \noindent{\it #1. }\nopagebreak}

\begin{document}
\newcommand{\Eq}[1]{Eq.~(\ref{#1})}
\newcommand{\Eqs}[1]{Eqs.~(\ref{#1})}
\newcommand{\eq}[1]{(\ref{#1})}
\newcommand{\ds}[1]{\displaystyle }
\newcommand{\bra}[1]{\left<#1\right|}
\newcommand{\ket}[1]{\left|#1\right>}
\newcommand{\braket}[2]{\left.\left<#1\right|#2\right>}
\newcommand{\blue}{\color{blue}}
\newcommand{\bea}{\begin{eqnarray}}
\newcommand{\eea}{\end{eqnarray}}
\newcommand{\be}{\begin{equation}}
\newcommand{\ee}{\end{equation}}
\newcommand{\red}{\color{red}}
\newcommand{\gray}{\color{gray}}
\newcommand{\black}{\color{black}}

\newcommand{\ca}[1]{{\cal #1}}
\newcommand{\sgn}{{\mathrm{sgn}}}
\newcommand{\rme}{{\mathrm{e}}}
\newcommand{\rmd}{{\mathrm{d}}} 
\newcommand{\nn}{\nonumber}
\newcommand{\E}{\epsilon}

\title{\sffamily\bfseries\large Large Orders  and Strong-Coupling Limit in Functional Renormalization}
\author{\sffamily\bfseries\normalsize   Mikhail N. Semeikin and Kay J\"org Wiese}
\affiliation{CNRS-Laboratoire de Physique de l'Ecole Normale Sup\'erieure, PSL, ENS, Sorbonne Universit\'e, Universit\'e Paris Cit\'e, 24 rue Lhomond, 75005 Paris, France}

\begin{abstract}
We study the  large-order behavior of the functional renormalization group (FRG). 
For a  model in dimension zero, we establish Borel-summability for a large class of microscopic couplings. 
Writing the derivatives of FRG as  contour integrals, we   express the 
Borel-transform as well as the original series   as     integrals. Taking the  strong-coupling limit  in this 
 representation, we  show that all short-ranged microscopic disorders flow to the same universal fixed point. 
Our results are relevant for FRG in disordered elastic systems.
\end{abstract}

\maketitle

\section{Introduction}
Perturbative expansions are a   work horse in theo\-retical physics. Most of them   
       are not converging, but asymptotic series \cite{Dyson1952,Lipatov1977a,Zinn-Justin1981,LeGuillouZinn-Justin1990,DavidWiese2004,KleinertSchulte-FrohlindeBook}. The main strategy to obtain a series with a  finite radius of convergence  is 
to define its Borel transform by dividing  its $n$-th series coefficient by $n!$.
One then  continues the latter  and   reconstructs the original function  via an integral over this analytic continuation.  The aim is to extend the range of applicability from   small expansion parameters, where the series naively converges, to  larger ones. Techniques using Pad\'e-Borel resummation or conformal mappings are successful here \cite{Zinn-Justin1981,LeGuillouZinn-Justin1990,KompanietsPanzer2017,KompanietsWiese2019,Kompaniets2016,DavidWiese2004}, and were employed for the $\epsilon$-expansion   of perturbative RG \cite{Kompaniets2016,KompanietsPanzer2017,BatkovichChetyrkinKompaniets2016,KleinertSchulte-FrohlindeBook}.
In simpler examples, as the anharmonic oscillator \cite{BenderWu1969}, one can go further, and use {\em resurgence} \cite{Dorigoni2019,AnicetoBasarSchiappa2019,MarinoLectureNotes} to 
reach finite couplings. Borel resummation    identifies  singularities of the 
Borel transform, and using this information   extends the domain of convergence. It is not effective in  reaching strong coupling. 

An additional   problem arises when the  microscopic set of couplings is itself  a function,  as in the functional renormalization group (FRG) treatment of disordered systems. 
 In   FRG, one uses a confining potential of strength $m^2$ to obtain the effective disorder in terms of the bare one, order by order in 
$\lambda = m^{d-4}$. Varying the FRG scale $m$ allows one to   obtain an {\em approximate} solution 
in the limit of $m \to 0$, 
i.e.~$\lambda \to \infty$.  

Here we consider a specific model in dimension $d=0$, which is later derived from the field theory of disordered elastic manifolds (for a review, see \cite{Wiese2021}), 
in which we can take the limit  of $\lambda \to \infty$ directly.
We wish to answer the following fundamental questions:  What is the large-order behavior of FRG?
Is it   Borel-summable? 
How can we study its strong-coupling limit? And how does universality arise?

\section{Setting the stage}
In order to address these questions, we   start with  the $O(2)$ model on a single site. This is not only the simplest  possible  model,
but key formulas will prove useful later. 
Consider the partition function,    
\be\label{1}
\ca Z_{O(2)}(\lambda) :=  \int_{  \tilde \phi, \phi}   \rme^{- \tilde \phi \phi - \lambda  \tilde \phi^2 \phi^2} \ , \quad \ca Z(0) = 1.
\ee
Here $\phi$ and $\tilde \phi$ are complex conjugate fields. 
Analysis proceeds via Wick's theorem, using the   measure induced by $\rme^{-\tilde \phi \phi}$,
\be\label{Wick}
\left< \tilde \phi^n f(\phi)\right>= (\partial_{\phi})^n 
f(\phi)\Big|_{\phi=0}
\quad \Rightarrow \quad
\left< \tilde \phi^n \phi^m\right>_0 =  n!\, \delta _{n,m}  .
\ee
This implies that 
\be
\ca Z_{O(2)}(\lambda) = \sum_{n=0}^\infty \frac{(2n)!}{n!} (-\lambda)^n.
\ee
Stirling's formula shows that this series is divergent. Its {\em Borel transform}, obtained by dividing the $n$-th series coefficient by $n!$, has a finite radius of convergence,
\be\label{4}
\ca Z_{O(2)}^{\rm B}(t) := \sum_{n=0}^\infty \frac{(2n)!}{(n!)^2} (-t)^n = \frac{1}{\sqrt{1+4t}}.
\ee
$\ca Z_{O(2)}^{\rm B}(t)$ has a branch cut   starting at $t=-1/4$, and  its {\em analytic continuation} is well defined for   $t>0$. 
This allows one to obtain $\ca Z_{O(2)}(\lambda)$ via an {\em inverse Borel transform}
\be\label{5}
\ca Z_{O(2)}(\lambda) = \int_0^\infty \rmd t \, \rme^{-t } \ca Z_{O(2)}^{\rm B}(t \lambda) = \sqrt {\frac{\pi }{4 \lambda}} \, \rme^{ \frac{1}{4 \lambda}} \text{erfc}\Big(\frac{1}{2
   \sqrt{\lambda}}\Big).
\ee
The crucial step in this resummation is our ability to analytically continue the Borel-transform $\ca Z_{O(2)}^{\rm B}(t)$ beyond its radius of convergence of $1/4$, to the   positive real  axis.

When no analytic result is available, the standard procedure   is to do a saddle-point (instanton) analysis of the integral \cite{Dyson1952,Lipatov1977a,Zinn-Justin1981,LeGuillouZinn-Justin1990,DavidWiese2004}, and then use resummation techniques or   resurgence. In practice one is often constrained to either approximate $\ca Z_{\rm B}(t)$ via a Pad\'e approximant \cite{Zinn-Justin1981}, Meijer G-function \cite{MeraPedersenNikolic2018}, or use a conformal mapping \cite{Zinn-Justin1981,KompanietsPanzer2017,Kompaniets2016}. While this allows one to extend the range of convergence, say by a factor of five, the question of the strong-coupling behavior   remains elusive. 

\section{Resummation of a functional expansion}
Let us proceed to a 0-dimensional   model for functional RG (norm as in \Eq{1}), 
\be\label{6}
\ca Z_{\rm FRG} (w,\lambda) :=  \int_{  \tilde \phi,  \phi }   \rme^{- \tilde \phi (\phi-w) + \lambda \tilde \phi^2[\Delta (0) - \Delta(\phi) ]}  .
\ee
At this stage, this is a  mathematical problem;   we  show later   its  significance for depinning. We   assume that $\Delta(\phi)$ is an analytic function,  fast and monotonously 
decaying  for $\phi\ge 0$, and that $\Delta(0)=1$. A good example is $\Delta(\phi)= \rme^{-\phi}$. 
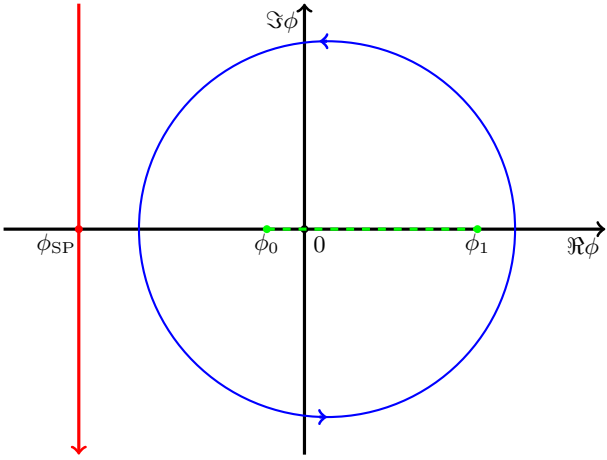
\begin{figure}[t]
{{\begin{tikzpicture}
\draw [->,very thick] (-4,0) -- (4,0);
\draw [->,very thick] (0,-3) -- (0,3);
\node (0) at  (0,0) {};
\node (0label) at  (.2,-.2) {$0$};
\node (m1) at  (0,2.75) {$\hspace{-6mm}\Im \phi$};
\node (m2) at  (4,-0.25) {$\hspace{-6mm}\Re \phi$};
\fill (0) circle (1.5pt);
\node (xSPlabel) at (-3.3,-0.2) {$ \phi_{\rm SP}$};
\coordinate (xSP) at  (-3,0);
\fill [red] (xSP) circle (1.5pt);
\node (x0label) at (-.5,-0.2) {$ \phi_{0}$};
\coordinate  (x0) at  (-.5,0) ;
\fill [green] (x0) circle (1.5pt);
\node (x1label) at (2.3,-0.2) {$ \phi_{1}$};
\coordinate  (x1) at  (2.3,0) ;
\fill [green] (x1) circle (1.5pt);
\draw [->,red,very thick] (-3,3) -- (-3,-3);
\draw [green,very thick,dashed] (x0) -- (x1);
\draw [blue,thick] (0.3,0) circle (2.5);
\draw [->,blue,very thick] (0.3,2.5) -- (0.2,2.5);
\draw [->,blue,very thick] (0.2,-2.5) -- (0.3,-2.5);
\end{tikzpicture}}}
\caption{The different paths and contour integrals. In blue the one used for \Eqs{15} and \eq{166}, encircling the cut in \Eq{C2B-final} (green/dashed). In red the path used for the derivation of \Eq{C-asymptotics2} which passes through $\phi_{\rm SP}$.}
\label{f:The different paths and contour integrals}
\end{figure}%
The field $\phi$  
has an expectation $w$.  Wick's theorem \eq{Wick}  allows us to write the perturbative expansion for $w>0$,  
\bea\label{118}
\ca Z_{\rm FRG} (w,\lambda) &:=& \sum_{n=0}^\infty \lambda^n  \ca Z_{\rm FRG}^{(n)}(w) , \\
  \ca Z_{\rm FRG}^{(n)}(w) &:=& \frac1{n! } (\partial_w)^{2n} \big[1- \Delta(w) \big]^n .
\label{118bis}
\eea
To  evaluate    \Eq{6}  non-perturbatively,   integration contours need to be specified. 
As we   show later, this is not an obvious task. 
Therefore we {\em define}  our model by 
\Eqs{118}--\eq{118bis}. The latter  are  motivated by perturbative results for the  renormalization of disordered elastic manifolds in dimension $d=0$ \cite{ChauveLeDoussalWiese2000a,LeDoussalWieseChauve2003,LeDoussalWieseChauve2002,WieseHusemannLeDoussal2018,HusemannWiese2017,SemeikinWiese2024}, for which $\Delta(\phi)$ is the microscopic disorder correlator.

Let us start with the large-order behavior of $\ca Z_{\rm FRG}^{(n)}(w)$. This  is given by the saddle point of \Eq{6} over both $\phi$ and $\tilde \phi$. It implies two saddle-point equations, is quite formal, and difficult to control. 
A more powerful approach is to evaluate    \Eq{118bis}    via the residue theorem, 
\bea\label{141}
\ca Z_{\rm FRG}^{(n)}(w) &=& {\frac{(2n)!}{n!  } \frac1{2\pi i} \oint \frac{\rmd \phi} \phi g_w(\phi)^n }, \\
  \quad g_w(\phi)&:=& \frac{1-\Delta(w{+}\phi)}{\phi^2}.
  \label{141bis}
\eea
The contour goes counter clockwise around the origin, see Fig.~\ref{f:The different paths and contour integrals}. 
It picks out the coefficient of order $\phi^0$ in the Laurent series at $\phi=0$.
 Since $\Delta(\phi )$ is bounded for $\Re \phi >0$, we can push the path in that domain to $\infty$. We expect a saddle point (SP) elsewhere, given by  
\be\label{SP}
\frac{\rmd}{\rmd \phi} g_w(\phi) = 0. 
\ee
To make our analysis concrete,   set $\Delta(\phi):=\rme^{-\phi}$. 
For $w=0$, the saddle point is at 
\bea
\phi_{\rm SP} &=& -W (- {2}\,{ \rme^{-2}} )-2 = -1.59362, \\
g_0(\phi_{\rm SP}) &=& -1.54414, 
\eea
where $W$ is the Lambert $W$ function. 
\begin{figure}
\Fig{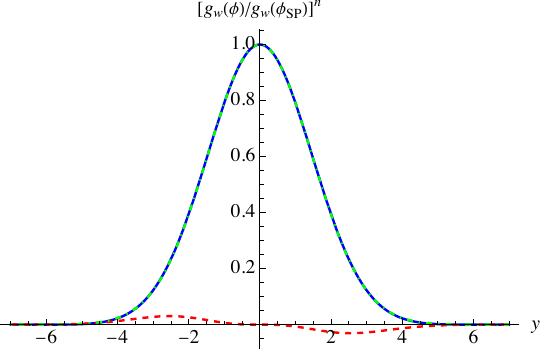}
\caption{Plot of $[g_w(\phi)/g_w(\phi_{\rm SP}) ]^n$ for $w=0$, $n=100$, with real part in blue (solid) and imaginary part in red (dashed);
$\phi= \phi_{\rm SP}+ i y/\sqrt{n}$, as indicated by the red curve on Fig.~\ref{f:The different paths and contour integrals}.
In green (dot-dashed) $\exp(-\frac{g_w''(\phi)}{     g_w(\phi)}\frac{y^2}2 )|_{\phi=\phi_{\rm SP}}$, whose integration leads to \Eq{C-asymptotics2}.} 
\label{f:SPintegral}
\end{figure}%
Fig.~\ref{f:SPintegral} shows that the large-order behavior of \Eq{118bis} is captured by the   integral running over $\phi = \phi_{\rm SP} + i \mathbb R$ (see Fig.~\ref{f:The different paths and contour integrals} for the path). 
This gives the leading order of the large-$n$ behavior, 
\be\label{C-asymptotics2}
{\ca Z_{\rm FRG}^{(n)}(w)   \simeq    \frac{ \Gamma(2n{+}1/2)}{\Gamma(n{+}1) \sqrt{\pi}}  {\left[ {g_w(\phi)} \right]^n}  
\sqrt{\frac{g_w(\phi)}{ g_w''(\phi)}}\Bigg|_{\phi=\phi_{\rm SP}}}.
\ee
The large-order behavior is asymptotic and its Borel transform exists, as $\Gamma(2n{+}1/2)/\Gamma(n{+}1) \simeq n!$.
The saddle point is at negative $\phi$, on the analytic continuation of the  branch for  $\phi\ge 0$,   outside its physically relevant domain.  
A numerical check  for  $n=100$ is shown in Fig.~\ref{f:SPintegral}. 
The relative error for $\ca Z_{\rm FRG}^{(n)}(0) $ is $10^{-4}$,
which can systematically be improved 
by further $1/n$ corrections. The latter are   relevant for resurgence \cite{SemeikinWieseLargeOrdersLongVersion}.

When changing the microscopic disorder from \(\Delta(\phi) = \rme^{-\phi}\) to \(\Delta(\phi) = \rme^{-\phi - a \phi^2}\), there is a critical \(a_{\rm c} \approx 0.0649\) beyond which 
the real saddle point at \(w=0\) disappears.   
It is replaced by an  infinity of pairs of complex saddle points,  corresponding  to 
a more intricate resurgent structure \cite{SemeikinWieseLargeOrdersLongVersion}.

\section{Borel transform}
  Define the Borel transform of \Eqs{118}-\eq{141} as 
\be\label{15}
\ca Z_{\rm FRG}^{\rm B}(w,t)   :=  \sum_{n=0}^\infty  \frac{(2n)!}{(n!)^2  }  \frac{t^n}{2\pi i} \oint \frac{\rmd \phi} \phi g_w(\phi)^n . 
\ee
Exchanging sum and integration,  \Eq{4} yields 
\be\label{166}
  { \ca Z_{\rm FRG}^{\rm B}(w,t) = \oint \frac{\rmd \phi }{2\pi i \phi} \frac1{\sqrt{1- 4 t g_w(\phi)}}}.
\ee
While \Eq{141} is valid for any contour circling the origin, in order to avoid the branch cut induced by  the denominator in \Eq{166}, 
 one needs to make the contour in \Eq{166} large enough, see Fig.~\ref{f:The different paths and contour integrals}.
One can then shrink the contour until it hugs the branch cut. Evaluating the discontinuity across the cut, we can  rewrite \Eq{166} as  
\be\label{C2B-final}
\ca Z_{\rm FRG}^{\rm B}(w,t)  =   \frac1\pi \int_{\phi_0}^{\phi_1} \rmd \phi \frac{1}{   \sqrt{4 t 
  [ 1{-}\Delta(w{+}\phi)]-\phi^2}},
\ee
where $\phi_0 \le 0 <\phi_1$ are the two zeros of the denominator, and the sign inside the square root is reversed between \Eqs{166} and \eq{C2B-final}.
For $w=0$, $\phi_0=0$. 
One could extend this integral from $-\infty $ to $\infty$, if one keeps only the real part of the integrand. A numerical check of \Eqs{166} and 
\eq{C2B-final} is presented in 
Fig.~\ref{ZBorel-lambda-3ways-exp} of the appendix.

\section{Inverse Borel transform}
Using \Eq{5},  the inverse Borel transform (from $t$ to $\lambda$) of the integrand in \Eq{166} is  (noting $g:=g_w(\phi)$)
\bea\label{159}
&&\int_0^{\infty } \frac{\rme^{-t}}{\sqrt{1- 4 \lambda gt }} \, \rmd t = \frac{\sqrt{\pi } \rme^{-\frac{1}{4 \lambda 
   g}}}{2} \frac{ \text{erfc}\Big(\frac{1}{2
   \sqrt{- \lambda  g}}\Big)}{
   \sqrt{-\lambda g}} \nn\\
&&   = \frac{\sqrt{\pi } \rme^{-\frac{1}{4 \lambda 
   g}}}{2}\bigg [ \frac1{\sqrt{-\lambda g}}+ \sum_{n\in \mathbb N} \frac{a_n}{(g\lambda)^n} \bigg].  
\eea
On the second line is the large-$\lambda$ expansion.
The key observation is that the terms $\sim a_n$  are analytic in   $\phi$  around  the origin, and thus do not contribute to the integral \eq{166}. As a consequence, 
the latter can   be  simplified to
\be\label{20}
 \ca Z_{\rm FRG}(w,\lambda) = \oint \frac{\rmd \phi }{2\pi i \phi}  \frac{\sqrt{\pi } \rme^{-\frac{1}{4 \lambda 
   g_w(\phi)}}}{2 \sqrt{-\lambda g_w(\phi)}}  .
\ee   
In order for this equality to be valid, the contour is not allowed to cross the cut which now extends to $\phi=\infty$, and  
 starts at $\phi=-w$. 
As in the derivation of \Eq{C2B-final}, we can simplify \Eq{20} by retaining only the discontinuity across the cut, 
\be \label{C2final}  
 \ca Z_{\rm FRG}(w,\lambda) = 
    {\frac{1}{\sqrt{4\pi\lambda} }}\int_0^\infty \rmd \phi\,   \frac{\rme^{\frac{-(\phi-w)^2 }{4 \lambda[1-\Delta (\phi)]}} }{\sqrt{1-\Delta(\phi)}}.
\ee
To arrive here, we   moved the factor of $1/\phi$ inside the square root, evaluated its discontinuity,  and finally shifted $\phi\to \phi+w$. This result is checked on Fig.~\ref{Delta2ofw-NEW-lambda=10} of the appendix.
Finally,   \Eq{C2final} can be derived from \Eq{6}, if one choses for   the integration contours    $\tilde \phi \in i \mathbb R$, and  $\phi\ge 0$. 

\section{Strong-coupling behavior}
\Eq{C2final} allows us to extract the large-$\lambda$ behavior. The key observation is that due to the factor of $1/\lambda$ in the exponent, larger and larger values for $\phi$ contribute. On these scales, $\Delta(\phi)$ is negligible and can be dropped, leading to 
\bea \label{C2asymp}  
 \ca Z_{\rm FRG}(w,\lambda) &\simeq&  
    {\frac{1}{\sqrt{4\pi\lambda} }}\int_0^\infty \rmd \phi\,    {\rme^{-\frac{(\phi-w)^2 }{4 \lambda}} } \nn\\
    &=& {\frac{1}{\sqrt{4\pi}  }}\int_0^\infty \rmd \phi\,    {\rme^{-\frac{(\phi-w/\sqrt{\lambda})^2 }{4 }} }.
\eea
The second line shows that the limit $ \ca Z_{\rm FRG}^\infty(w) :=\lim_{\lambda \to \infty} \ca Z_{\rm FRG}(w \sqrt{\lambda},\lambda)$ exists, and is given by 
\be
\ca Z_{\rm FRG}^\infty(w) = \frac{1}{2}\left[1+ 
   \text{erf}\Big(\frac{w}{2}\Big)\right].
\label{170}
\ee
\begin{figure}
\Fig{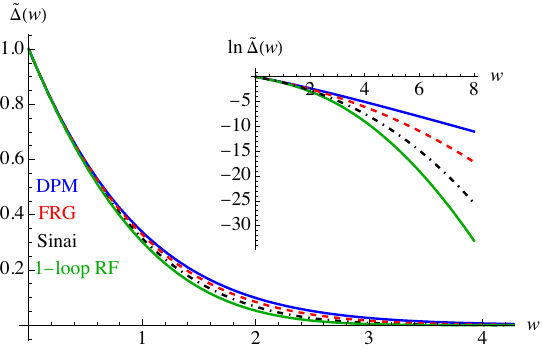}
\caption{Different solutions for $\tilde \Delta(w)$, all rescaled to $\tilde \Delta(0) =|\tilde \Delta'(0)|=1 $. From top to bottom:  driven particle (DPM) in Gaussian disorder (blue),  Eq.~(85) of \cite{LeDoussalWiese2008a}, \Eq{25} (red, dashed), Sinai model, Eq.~(202) of \cite{Wiese2021} (black, dot-dashed), and the 1-loop random-field fixed point,  Eq.~(88) of \cite{Wiese2021} (green, solid).}
\label{Delta-comparison}
\end{figure}%
To derive this it is essential that the singularity in the denominator of \Eq{C2final} is integrable.  
Numerically we   checked the passage from \Eq{C2final}  to \Eq{C2asymp} for $\lambda$ up to $10^{20}$. 

Finally, \Eqs{118} and \eq{118bis} imply that $ \ca Z_{\rm FRG}(w,\lambda) = 1- \lambda\Delta_{\rm FRG}''(w,\lambda)$. 
Therefore the dimensionless rescaled limit for   $\Delta''_{\rm FRG}$ reads
\be\label{24}
\tilde \Delta''_{\rm FRG}(w):= \lim_{\lambda \to \infty} \lambda^{-1}\Delta_{\rm FRG}''(w\sqrt{\lambda}, \lambda) =   
\frac12 \text{erfc}\Big(\frac{w}{2}\Big).
\ee
Integrating twice and using $\tilde \Delta_{\rm FRG}(\infty)=0$ yields
\be\label{25}
\tilde \Delta_{\rm FRG}(w)  = 
\frac{w^2+2}{4}
   \text{erfc}\left(\frac{w}{2}\right)-\frac{  \rme^{-\frac{w^2}{4}} w}{2\sqrt{\pi}}.
\ee
What is remarkable about \Eq{C2final} is that the final result,   given in \Eq{170}, is largely independent of the microscopic   $\Delta(\phi)$. 
What we used   is that $\Delta(\phi)$ is analytic,  has a linear cusp at the origin, and decays quickly. The cusp is a technical requirement, necessary to transform 
the contour integral into a cut integral. We believe that this is more a technical constraint than a physical one: we could regularize  the 
microscopic disorder    to obtain a linear cusp, and then remove the regularization. 
We have studied this for   $\Delta(\phi)= \rme^{-\phi^2}$. While we clearly see that convergence is non-uniform and slow, we have no indication 
that the process does not converge, or converges against a different fixed point. On the practical side, when applied to disordered systems,    as 
 the  disorder  usually lives on a grid,   we can well approximate it by a function with a linear cusp. 

What is reassuring about our findings  is that while it is believed that all microscopic disorders converge to 
the same FRG fixed point, this has only be seen perturbatively \cite{ChauveLeDoussalWiese2000a,LeDoussalWieseChauve2003,LeDoussalWieseChauve2002,WieseHusemannLeDoussal2018,HusemannWiese2017,SemeikinWiese2024}, in simulations \cite{MiddletonLeDoussalWiese2006,RossoLeDoussalWiese2006a,terBurgWiese2020} and in experiments \cite{WieseBercyMelkonyanBizebard2019,terBurgBohnDurinSommerWiese2021,terBurgRissoneRicoPastoRitortWiese2023}.  The mechanism by 
which this happens here is non-perturbative, and apparently   robust.  

Finally, let us compare the shape of $\tilde \Delta(w)$ as derived in \Eq{25} to other analytical solutions (Fig.~\ref{Delta-comparison}): A $d=0$ solution for depinning, the $d=0$ solution in equilibrium with random-field (RF) disorder (Sinai model), and the 1-loop solution in the RF universality class. 
While   these solutions are similar, they are   distinct and allow one to determine the universality class, as was done 
for magnetic domain walls   \cite{terBurgBohnDurinSommerWiese2021}.

\enlargethispage{1.5mm}

\section{Field theory for disordered elastic systems}
Let us   connect our findings to the field theory of disordered elastic systems. 
This is best done by comparing to the formulation of Ref.~\cite{WieseFedorenko2018,WieseFedorenko2019} which uses  Grassmanian variables (``{\em supersymmetry}'') \cite{ParisiSourlas1979,ParisiSourlas1981,BrydgesImbrieSlade2009,BrydgesImbrie2003} to average over disorder. The relevant action contains two physical replicas located at positions $u_1$ and $u_2$. 
Denoting their center of mass by $u$, and their difference by $\phi$,   
only $\phi$ appears inside the disorder correlator $\Delta$, and    $u$ decouples. The corresponding action becomes
(see  the appendix)
\begin{align}
\label{FRG-SUSY-ACTION} {\cal S}  =   
\parbox[c][0mm][c]{3ex}{$\displaystyle \int_{\ensuremath{x}}$} \rule{0mm}{3ex}
&\tilde \phi(x) (m^2-\nabla^2)[\phi(x)-w]  
\nn\\ &
+  \sum_{a=1}^{{2}}\bar \psi_{a} (x)
(m^2-\nabla^2)\psi_{a} (x)\nn\\
&+  \tilde \phi(x)^2 \Big[ \Delta\big(\phi(x)\big)-\Delta(0)\Big] 
\nn\\
& +  \tilde \phi(x) \Delta'\big(\phi(x)\big)\Big[\bar \psi_{2}(x) \psi_{2}(x)+\bar \psi_{1}(x) \psi_{1}(x)\Big]
\nn\\ &
+ \bar \psi_{2}(x) \psi_{2}(x) \bar \psi_{1}(x) \psi_{1}(x) \Delta''\big(\phi(x)\big)\ .
\end{align}
Here $\tilde \phi$ and $\phi$ are bosonic fields (complex numbers), while $\bar \psi_i$ and $\psi_i$ are Grassmann variables \cite{Brezin1985}. 
To understand this action, let us temporarily drop the fermionic fields. Taking   dimension $d\to 0$, and rescaling $\tilde \phi\to \tilde \phi/m^2$,  
we get   the 
 model of \Eqs{6}-\eq{118bis}, 
\bea\label{28}
 \ca Z_{\rm FRG}(w,\lambda) &\equiv& \ca Z^{\ca S}_{\rm bos}(w,\lambda )\big|_{d=0}:= \int_{\phi, \tilde \phi} \rme^{-\ca S|_{\psi_i \to 0}},\qquad \\
 \lambda &\equiv&  {m^{-4}}. \label{lamda*m4=1}
\eea
\Eq{lamda*m4=1} implies that $w\sim \sqrt\lambda =  m^{-2}$, thus the scaling exponent of the field $\phi$,  a.k.a.\ the roughness exponent $\zeta$, is 
\be
\zeta = 2, 
\ee
which   also holds for the  action \eq{FRG-SUSY-ACTION}. 
By construction, the   partition function   of the latter  over all bosonic and Grassmann fields is $1$, 
\be\label{29}
\ca Z_{\ca S}(w,\lambda ):= \left< 1 \right>_{\ca S}=1, \quad \left< \ca O\right>_{\ca S}:= \int_{\phi, \tilde \phi,\bar \psi_1,\psi_1,\bar \psi_2,\psi_2} \rme^{-\ca S} \ca O.
\ee
The renormalized $\Delta(w)$ is  given \cite{Wiese2021} by  the connected expectation of $m^4(\phi-w)^2/2$, 
\be\label{30}
\Delta_{\rm Susy}(0,\lambda) - \Delta_{\rm Susy}(w,\lambda) = \frac{m^4}2 \left<  {(\phi{-}w)^2}  \right>^{\rm c}_{\ca S}. 
\ee
\noindent
This function has a limit, 
\be\label{32}
\tilde \Delta_{\rm Susy}(w) : =   \lim_{m\to 0} \Delta_{\rm Susy}(w m^{-2},m^{-4}) .
\ee
It is non-trivial to show that the functions  defined in \Eqs{32} and \eq{25} agree (see the appendix and \cite{SemeikinWieseLargeOrdersLongVersion}),
\be\label{central}
\tilde \Delta_{\rm Susy}(w) = \tilde \Delta_{\rm FRG}(w). 
\ee
Thus what we   obtained for the simple model   \eq{6} also applies to the disordered system defined by the action \eq{FRG-SUSY-ACTION}.

\section{Applications}
Our results agree up to 1-loop order with that for  disordered elastic manifolds in equilibrium and at depinning \cite{ChauveLeDoussalWiese2000a,LeDoussalWieseChauve2002,LeDoussalWieseChauve2003,WieseHusemannLeDoussal2018,HusemannWiese2017,SemeikinWiese2024}. Beyond that, amplitudes are different in the $\epsilon$-expansion, and there are additional {\em anomalous terms} which are    hard to recuperate   \cite{SemeikinWieseLargeOrdersLongVersion}. 
While our   model  can formally be derived from a field theory in equilibrium, we do not believe $  \tilde \Delta_{\rm FRG}(w)$ to be relevant for a specific 
physical situation, even though the predicted roughness exponent  is equal to that of depinning, and the shape of  $\tilde \Delta_{\rm FRG}(w)$ on Fig.~\ref{Delta-comparison} is between   a driven particle and Sinai's model, both relevant in $d=0$.  

Given these caveats, we turn to the strengths of our approach.
Our   model contains all   ingredients of functional renormalization:   it shows that  the perturbative series  is Borel summable, how the limit of strong coupling is reached, that it cannot be inferred from   the large-order behavior, and how universality emerges. 
By connecting   to a   formulation  via superfields, we establish the connection to field theory.
 The zero-dimensional limit   is relevant for DNA unzipping \cite{terBurgRissoneRicoPastoRitortWiese2023} and  RNA/DNA peeling \cite{WieseBercyMelkonyanBizebard2019}, providing a concrete physical application. 
Since this limit   retains all relevant physics, as e.g.~avalanches, and can be assessed in an expansion in $\epsilon=4-d$,  the 
model \eq{118}-\eq{118bis}  is key  in understanding  functional RG and its $\epsilon$-expansion; 
as a solution of   the  model \eq{1}, even though somehow trivial, is key in understanding     the $\epsilon$-expansion in $\phi^4$-theory.

  Our work poses a solid  framework for the strong-coupling behavior in functional renormalization,   constraining the large-order and strong-coupling behavior in dimension $d>0$. 
We also saw that   to define the path integral non-perturbatively, one needs to specify the integration contours, and   constrain   variables to part of their physically allowed domains.   This     restricts   theories in dimension $d>0$.

\appendix

\medskip


\centerline{\bf Appendices}
\smallskip

\section{Field Theory} 
Building on the Susy formulation of \cite{Wiese2004},   Ref.~\cite{WieseFedorenko2018}  introduces two physical copies located at $u_1$ and $u_2$, which are subject to confining potentials displaced by $w$, s.t.\ their difference $\phi:=u_1-u_2$ has  expectation $\left<\phi\right>=w$; its  center of mass is $u:=(u_1+u_2)/2$. 
The field theory, given in   Eq.~(36)  of \cite{WieseFedorenko2018} reads
\begin{align}
\label{app1} {\cal S} 
=   \parbox[c][0mm][c]{3ex}{$\displaystyle \int_{\ensuremath{x}}$}\rule{0mm}{3ex} 
 &\tilde \phi(x) (m^2-\nabla^2)[\phi(x)-w]+ \tilde  u(x) (m^2-\nabla^2) u(x)
\nn\\ &
+   \sum_{a=1}^2\bar \psi_{a} (x)
(m^2-\nabla^2)\psi_{a} (x)\nn\\
&+  \tilde \phi(x)^2 \Big[ \Delta\big(\phi(x)\big)-\Delta(0)\Big] 
\nn\\
&-\frac14 \tilde u(x)^2\Big[ \Delta\big(\phi(x)\big)+\Delta(0)\Big]
\nn\\ &
+ \frac12 \tilde u(x) \Delta'\big(\phi(x)\big)\Big[\bar \psi_{2}(x) \psi_{2}(x)-\bar \psi_{1}(x) \psi_{1}(x)\Big] \nn\\& +  \tilde \phi(x) \Delta'\big(\phi(x)\big)\Big[\bar \psi_{2}(x) \psi_{2}(x)+\bar \psi_{1}(x) \psi_{1}(x)\Big]
\nn\\ &
+ \bar \psi_{2}(x) \psi_{2}(x) \bar \psi_{1}(x) \psi_{1}(x) \Delta''\big(\phi(x)\big)\ .
 \end{align}
Here $\tilde u$ and $\tilde \phi$ are the response fields for $u$ and $\phi$, while $\psi_i$ and $\bar \psi_i$ are Grassmannian variables introduced to ensure that the partition function equals 1.  
Integrating over $u$ forces $\tilde u \to 0$, resulting in \Eq{FRG-SUSY-ACTION} of the main text. 
\begin{widetext}
\noindent Let us next take dimension $d=0$ in action \eq{FRG-SUSY-ACTION}, and   integrate over the Grassmann variables. This gives the partition function 
\bea\label{34}
\ca Z &=& \frac1{m^4} \int_{\phi}\int_{\tilde \phi}\left\{\Big[\tilde \phi \Delta '(\phi)+m^2\Big]^2{-}\Delta ''(\phi )\right\}
\exp \!\bigg(-\left[\tilde \phi ^2
   \big(\Delta (\phi )-\Delta (0)\big)\right]-m^2\tilde \phi (\phi -w)\bigg). \qquad 
\eea
\noindent Integrating   $\tilde \phi$ over the imaginary axis yields
\be\label{app-cut-integral-Z}
\ca Z   =\frac{1}{2 m^2\sqrt{\pi}}\int_0^\infty \rmd \phi \,\bigg\{ m^4-\Delta ''(\phi )+\frac{m^4 (w-\phi)^2 \Delta '(\phi )^2}{4 [\Delta(0)-\Delta (\phi)]^2}
-\frac{\Delta '(\phi )   \left[\Delta '(\phi )+2 m^4 (w-\phi )\right]}{2 [\Delta(0)-\Delta (\phi )]}\bigg\} 
\frac{\rme^{-\frac{m^4 (w-\phi   )^2}{4 [\Delta (0)-\Delta (\phi   )]}}}{\sqrt{\Delta (0)-\Delta   (\phi )}}.
\ee
\end{widetext}
After \Eq{C2final} we stated that the latter  can be derived from \Eq{6} if $\tilde \phi \in i \mathbb R$ and $\phi>0$.  
We use the same  prescription  to pass from \Eq{34} to \Eq{app-cut-integral-Z},
checking that $\ca Z=1$ in  \Eq{app-cut-integral-Z}. We then evaluated \Eqs{30}-\eq{32}  perturbatively and numerically, proving \Eq{central} (for details see \cite{SemeikinWieseLargeOrdersLongVersion}).

\section{Additional numerical checks}
\begin{figure}[b]
\Fig{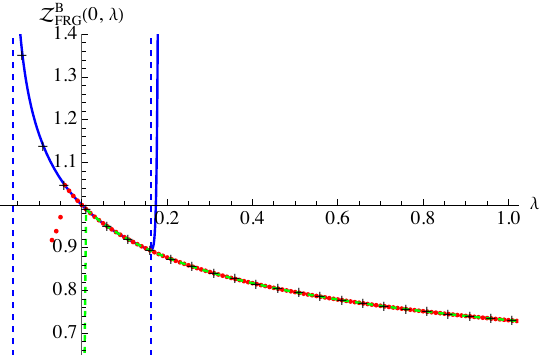}
\caption{$\ca Z_{\rm FRG}^{\rm B}(w=0,\lambda)$, evaluated in four different ways: (i) explicit sum from derivatives as given in \Eqs{118} and \eq{118bis} (blue solid line). 
The vertical blue-dashed lines indicate its  radius of convergence estimated from \Eq{C-asymptotics2}.  (ii)
the contour integral \eq{166} (red dots),  (iii) the cut integral \eq{C2B-final} (green dashed), and (iv) a diagonal Pad\'e resummation of the original series (black crosses).
Both integral representations work for   $\lambda$ larger than the radius of convergence of the series (but are as expected problematic for negative $\lambda$). 
}
\label{ZBorel-lambda-3ways-exp}
\end{figure}%
Fig.~\ref{ZBorel-lambda-3ways-exp} shows   that for $w=0$, 
\be\label{37}
\ca Z_{\rm FRG}^{\rm B}(w) := \sum_{n=0}^\infty \frac{\lambda^n}{n!}  \ca Z_{\rm FRG}^{(n)}(w) , 
\ee
with $\ca Z_{\rm FRG}^{(n)}(w)$  defined in \Eq{118bis}, agrees with both \Eq{166} and \Eq{C2B-final}   inside its radius of convergence, at least for $\lambda>0$.
For $\lambda>0$ and outside the radius of convergence, the latter two agree with each other and a Pad\'e resummation of \
\Eq{37}. 

\begin{figure}[b]
\Fig{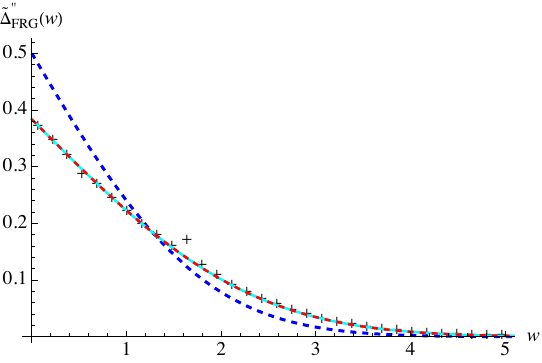}
\caption{The function $\tilde \Delta_{\rm FRG}''(w,\lambda):= 1-\ca Z_{\rm FRG}(w \sqrt{\lambda},\lambda)$ for $\lambda=10$, evaluated via: Pad\'e-Borel (black crosses) on the combinatorial series at order 100.
(Some glitches appear, and  Pad\'e-Borel breaks down for larger  $\lambda$; each Pad\'e is constructed at fixed $w$.) Evaluation of the integral \eq{C2final}  (cyan, solid), indistinguishable form 
an implementation which keeps the erfc of \Eq{159}  (red, dashed). In blue dashed the asymptotic form \eq{24}.}
\label{Delta2ofw-NEW-lambda=10}
\end{figure}

Fig.~\ref{Delta2ofw-NEW-lambda=10} shows the rescaled $\tilde \Delta''_{\rm FRG}(w)$ for $\lambda=10$, i.e.\ well outside the range of convergence of the Borel transform. We tested the   integral  \eq{20} against a Pad\'e-Borel approximation of the original series. Deviations for some values of $w$ are visible due to the large value of $\lambda$, but are absent for smaller $\lambda$ (not shown). We also tested that there is no difference when keeping the $\mbox{erfc}$ in \Eq{159}, instead of replacing it by $1$, as was done in the derivation of \Eq{20}. Finally,  the solution approaches the asymptotic form \eq{24}. We checked this convergence   for   $\lambda$ up to $10^{20}$ using the cut integral \eq{C2final} (not shown).

\smallskip

\acknowledgements
We are grateful to Andrei Fedorenko for stimulating discussions and many deep questions. 
We   profited from exchanges with Costas Bachas and Edouard Br\'ezin,  and feedback  from the referees.


\ifx\doi\undefined
\providecommand{\doi}[2]{\href{http://dx.doi.org/#1}{#2}}
\else
\renewcommand{\doi}[2]{\href{http://dx.doi.org/#1}{#2}}
\fi
\providecommand{\link}[2]{\href{#1}{#2}}
\providecommand{\arxiv}[1]{\href{http://arxiv.org/abs/#1}{#1}}
\providecommand{\hal}[1]{\href{https://hal.archives-ouvertes.fr/hal-#1}{hal-#1}}
\providecommand{\mrnumber}[1]{\href{https://mathscinet.ams.org/mathscinet/search/publdoc.html?pg1=MR&s1=#1&loc=fromreflist}{MR#1}}

\end{document}